\documentclass[12pt]{article}
\usepackage{fullpage,epsfig,graphics,amsbsy,amssymb,cancel,bbm}
\usepackage{psfrag,hyperref,color,slashed}
\usepackage{graphicx}
\usepackage{amsfonts,mathdots,dsfont}
\usepackage{amssymb,amsmath,amscd,mathrsfs}
\usepackage{tikz}
\usetikzlibrary{arrows,chains,matrix,positioning,scopes,snakes}

\makeatletter
\tikzset{join/.code=\tikzset{after node path={%
\ifx\tikzchainprevious\pgfutil@empty\else(\tikzchainprevious)%
edge[every join]#1(\tikzchaincurrent)\fi}}}
\makeatother

\tikzset{>=stealth',every on chain/.append style={join},
         every join/.style={->}}
\tikzset{
    >=stealth',
    punkt/.style={
           rectangle,
           rounded corners,
           draw=black, very thick,
           text width=6.5em,
           minimum height=2em,
           text centered},
    pil/.style={
           ->,
           thick,
           shorten <=2pt,
           shorten >=2pt,}
}

\newcommand{\BB}{\mathbb}

\newcommand{\FR}{\mathfrak}

\def\L{{\cal L}}

\newcommand{\bea}{\begin{eqnarray}}
\newcommand{\eea}{\end{eqnarray}}
\newcommand{\nn}{\nonumber}
\newcommand{\Tr}{\textrm{Tr}}

\newcommand{\im}{\textrm{Im}\,}
\newcommand{\re}{\textrm{Re}\,}
\newcommand{\To}{\Rightarrow}

\newcommand{\reeb}{\textrm{\scriptsize{$R$}}}
\newcommand{\sreeb}{\textrm{\tiny{$R$}}}
\newcommand{\gYM}{g_{\textrm{\tiny{$YM$}}}}

\def\gb{\beta}

\def\gd{\delta}
\def\ep{\epsilon}
\def\gt{\theta}

\def\gs{\sigma}

\def\gk{\kappa}
\def\gl{\lambda}
\def\Gl{\Lambda}

\def\go{\omega}

\DeclareMathAlphabet{\mathpzc}{OT1}{pzc}{m}{it}

\newcommand{\qed}{\nobreak \ifvmode \relax \else
      \ifdim\lastskip<1.5em \hskip-\lastskip
      \hskip1.5em plus0em minus0.5em \fi \nobreak
      \vrule height0.5em width0.5em depth0.00em\fi}

\begin{document}
\renewcommand{\theequation}{\thesection.\arabic{equation}}
\setcounter{page}{0}

\thispagestyle{empty}
\begin{flushright} \small
UUITP-18/13
 \end{flushright}
\smallskip
\begin{center} \LARGE
{\bf Factorization of 5D super Yang-Mills  \\
 on $Y^{p,q}$ spaces}
 \\[12mm] \normalsize
{\bf  Jian Qiu$^a$ and Maxim Zabzine$^b$} \\[8mm]
 {\small\it
${}^a$Math\'ematiques, Universit\'e du Luxembourg,\\
 Campus Kirchberg, G 106,  L-1359 Luxembourg\\
      \vspace{.5cm}
${}^b$Department of Physics and Astronomy,
     Uppsala university,\\
     Box 516,
     SE-75120 Uppsala,
     Sweden\\
   }
\end{center}
\vspace{7mm}
\begin{abstract}
 \noindent
   We continue our study on the partition function for 5D supersymmetric Yang-Mills theory on toric Sasaki-Einstein $Y^{p,q}$ manifolds.
   Previously, using the localisation technique we have computed the perturbative part of the partition function.
    In this work we show how the perturbative part factorises into four pieces, each corresponding to the perturbative
     answer of the same theory on $\mathbb{R}^4 \times S^1$.  This allows us to identify the equivariant parameters and to conjecture the full partition
      functions (including the instanton contributions) for $Y^{p,q}$ spaces.
      The conjectured partition function receives contributions only from singular contact instantons supported along the closed Reeb orbits.
         At the moment we are not able to prove this fact from the first principles.
    \end{abstract}

\eject
\normalsize
\tableofcontents
\section{Introduction}\label{sec_intro}

Starting from Pestun's work \cite{Pestun:2007rz} the localisation techniques have been widely used in calculating the partition functions and the expectation values of supersymmetric Wilson loops for a large class of supersymmetric gauge theories defined
  over compact manifolds. However, the main examples of the compact manifolds on which calculations have been carried out are
   the spheres $S^d$ and $S^{d-1} \times S^1$. In order to define the supersymmetric theory on curved space, we need that
    there should exist solutions of the Killing spinor equation or its different modifications related to the coupling to additional background fields.
    In odd dimensions and with the simplest setup, we need the manifold to be a simply connected
    Sasaki-Einstein manifold so as to accommodate some fraction of supersymmetry. In three dimensions the round $S^3$ is the unique
     simply connected Sasaki-Einstein manifold. While in five dimensions we have infinite families of such manifolds, amongst which the round $S^5$ is the only one that allows 8 supercharges.
     Besides $S^5$, there is also the family called the $Y^{p,q}$-manifolds and its generalisation the $L^{a,b, c}$-manifolds, both having the topology of $S^3 \times S^2$ and admitting a quarter supersymmetry.

 In \cite{Qiu:2013pta} the full perturbative partition function has been calculated for $Y^{p,q}$ spaces.  In this paper we continue to study
  the  partition function for $Y^{p,q}$ spaces. We prove the factorisation of the perturbative part into four pieces corresponding to
   four copies of pertubative partition function on $\mathbb{R}^4 \times S^1$.  This decomposition allows us to identify the equivariant parameters
    and conjecture the full partition function with instantons for the $Y^{p,q}$ spaces. The main novelty of the present study is that
     now we have exact results for an infinite number of compact manifolds in addition to $S^5$.  The factorisation properties of 3D and 5D supersymmetric Yang-Mills theory have been studied intensely also in \cite{Beem:2012mb,Nieri:2013yra,Nieri:2013vba}, our work in this direction provides more specimens and should bring a better understanding
      of the structure of the partition function of these theories.

  Let us now explain schematically the main result of this work.
We are mostly interested in a subclass of 5D metric contact manifolds, the toric Sasaki-Einstein manifolds. We recall the definition here in order to state our main result. The reader may consult \cite{2010arXiv1004.2461S} for a more pedagogical exposition.

We say a 5D manifold $M$ with metric $g_M$ is \emph{Sasaki} if its metric cone $C(M)$ with metric $\tilde g=dr^2+r^2 g_M$, $r\gneq 0$ is a K\"ahler manifold with complex structure denoted as $J$. The vector field $\ep=r\partial_r$ that generates the scaling in $r$ is called the homothetic vector field. The K\"ahler structure on $C(M)$ leads to a \emph{metric contact structure} on $M$, in particular, the Reeb vector and contact 1-form on $M$ are given by
\bea
\reeb=J(r\partial_r)~,~~~\gk=i(\bar\partial-\partial)\log r~,\nn
\eea
with the consequence that $g_M\reeb=\gk$ and that $\reeb$ is a Killing vector field.
The manifold $M$ is called \emph{toric} if there is
an effective, holomorphic and Hamiltonian action of the torus $T^3$ on the
corresponding K\"ahler cone $C(M)$, and the Reeb vector field is a linear combination of the torus action. And also $M$ is Sasaki-Einstein (SE) if it is Sasaki and its Ricci tensor satisfies $R_{mn}=4g_{mn}$. Our main examples $S^5$, $Y^{p,q}$-space discovered in \cite{Gauntlett:2004yd} and $L^{a,b,c}$-space discovered in \cite{Cvetic:2005ft} are all toric SE manifolds.

Let $e_a,~a=1,2,3$ be the vector field of the aforementioned $T^3$-action, with $\mu_a$ being the moment map. Because of the cone structure, the image of the moment map of the $T^3$-action is a polytope cone. For example, in the case of $S^5$, the cone has a triangle as its base, while for the case of $Y^{p,q}$ and $L^{a,b,c}$, the base is a quadrangle. Focusing on the quadrangle case, we denote by $\vec v_i,~i=1\cdots 4$ the four primitive normal vectors of the four faces of the polytope cone (the over arrow $\vec {}$ on $v_i$ will be kept only when it is conducive to the clarity). Here primitiveness means that $v_i=\sum_{a=1}^3 n_ae_a$, such that $n_a\in\BB{Z}$ and $\gcd(n_1, n_2, n_3)=1$. By assumption, the Reeb is given by $\reeb=\sum_{a=1}^3\reeb_ae_a$, then the plane $\sum_{a=1}^3\mu_a\reeb_a=1/2$ will intersect the cone at a polygon (which is the base of the cone) iff $\reeb$ can be written as
\bea \reeb=\sum_{i=1}^4\gl_iv_i~,~~~\gl_i>0~.\label{dual_cone_intro}\eea
We call this the \emph{dual cone} condition \cite{Martelli:2005tp}.
Thus $Y^{p,q}$ (or $L^{a,b,c}$) can be thought of as a $T^3$ fibration over the base quadrangle, a particular instance of such a base is given in Figure \ref{fig_toric_cone}. Similarly $S^5$ can be thought of a $T^3$ fibration over the base triangle.

Our main conjecture is that the full partition function for $Y^{p,q}$ space has the structure
\bea
&&\int\limits_{\FR{t}}da~e^{-\frac{8\pi^3 r\varrho}{\gYM^2}\,\Tr[a^2] }
Z^{pert}_{Y^{p,q}} \times\prod_{i=1}^4 \Big [ Z_{\mathbb{R}^4 \times S^1}^{inst} \left (ia, i m + \Delta_i,\gb_i,\ep_i, \ep'_i \right ) \Big ]~, \label{fullans-Ypq_intro}
\eea
where one has a copy of Nekrasov instanton partition function $Z_{\mathbb{R}^4 \times S^1}^{inst}$ for each corner on the quadrangle.  Here $\gb$ corresponds to the circumference of the circle, while $\ep,\;\ep'$ are the equivariant parameter for the two rotations   on   $\BB{R}^4\sim\BB{C}\times\BB{C}$. This conjecture is inspired by the fact that the perturbative part $Z^{pert}_{Y^{p,q}}$ has a factorisation into four copies of the perturbative Nekrasov partition function, each of which corresponds to a corner of the quadrangle, see section \ref{s-Yspace}
 \bea
 Z^{pert}_{Y^{p,q}} = e^{B^{Y^{p,q}}} \prod_{i=1}^4 Z_{\mathbb{R}^4 \times S^1}^{pert} \left (ia, i m + \Delta_i ,\gb_i,\ep_i, \ep'_i \right )~.
 \eea
 The parameters $\gb_i,\;\ep_i\;\ep_i'$ can be read off from the toric data alone, which will be explained in subsection \ref{sec_Ioepftd}. As an example, we are at the corner corresponding to the intersection of face 1 and 2. Then $\beta$ can be read off from the determinant of the following matrix
\bea \frac{\gb}{2\pi}={\det}^{-1}[\reeb,v_1,v_2]~.\nn\eea
For $\ep,\;\ep'$, we let $\vec n$ be an integer-entry 3-vector such that $\det[\vec n,v_1,v_2]=1$ (the existence of $\vec n$  is a consequence of the fact that the metric cone is smooth, see also subsection \ref{sec_Ioepftd}), then
\bea
 \ep =\det([\reeb,v_2,\vec n])~,~~~\ep'=\det[v_1,\reeb,\vec n]~.\nn\eea
It is important to stress that the identification of parameters $\ep, \ep'$ is not unique, one may always add to $\ep,\;\ep'$ integer multiples of $2\pi\gb^{-1}$.

Finally, for the hyper-multiplet, the effective mass that one plugs into the flat space result is shifted from the bare mass by
\bea
\Delta_i = \frac{1}{2\det[v_1,v_2,v_3]}\big(\det[v_2,v_3,\reeb]+\det[v_3,v_1,\reeb]+\det[v_1,v_2,\reeb]\big)- \frac{1}{2} (\epsilon_i +\epsilon'_i) ~,\label{shift_toric}\eea
here one can write the shift by using any triple of $v_i$. The shift $\Delta_i$ in mass is not unique and it is not physical in any sense.
 For $S^5$ and $Y^{p,q}$ spaces we are able to choose the parameters in
 such way that $\Delta_i=0$. However we do not know if it is possible for any toric Sasaki-Einstein manifold. The actual physical combination is
  $\Delta_i + 1/2 (\epsilon_i + \epsilon_i')$ and this is the same for all $i=1,..., 4$.

It is also necessary to point out that, though our calculation has been performed on $S^5$ and $Y^{p,q}$ spaces, the above discussion is likely to extend also to the $L^{a,b,c}$-space. While for the vector multiplet, whose cohomological complex (see subsection \ref{subs-localS}) requires only a K-contact structure, we believe that there is a similar structure of factorisation for the partition function, and that the final answer can be read off from nothing but the toric diagram.

The paper is organised as follows: In section \ref{s-flat} we review the Nekrasov partition function for $\mathbb{R}^4
\times S^1$ and lay down the conventions for the rest of the paper.  Section \ref{s-sphere} is devoted to the discussion of partition function
 on $S^5$, where we summarise some known results from the literature. We discuss the cohomological complex and some details of the localisation.
 In section \ref{s-Yspace} we repeat the same procedure for $Y^{p,q}$ spaces. We discuss the decomposition of
 the perturbative result into four pieces and conjecture the full answer for the partition function.
 In  section \ref{s-instantons} we   explain how to read off the parameters in partition function from the toric data.
 Section \ref{s-summary} contains a summary and discussion of some open questions. The paper contains two appendices,
 in appendix \ref{A-sines} some known facts about multiple
   zeta, gamma and sine functions are collected. While appendix \ref{A-gen-sines} contains
    some original results on the generalisations of these functions inspired by the partition function on $Y^{p,q}$ space.

\section{Partition function on $\mathbb{R}^4 \times S^1$}\label{s-flat}

 Let us consider the five dimensional supersymmetric gauge theories with 8 supercharges defined on $\mathbb{R}^4 \times S^1$.
 The partition function for these theories has been studied  starting from  the works \cite{Nekrasov:1996cz, Lawrence:1997jr, Nekrasov:2002qd}.
   Let us consider the five dimensional theory of vector multiplet coupled to hypermultiplet in representation $\underline{R}$ defined on
 $\mathbb{R}^4 \times S^1$ where  the circle $S^1$ has circumference $\beta$ with the appropriate twisted periodicity conditions    imposed
  on   fields \cite{Nekrasov:2002qd}. Equivalently we can impose the periodic boundary conditions on fields and extra insertions of twisting operators.
 Thus we define the following index (partition function)
  \bea
  Z^{full}_{\mathbb{R}^4 \times S^1} = {\rm Tr}_{\cal H} \left ( (-1)^{2(j_L + j_R)}e^{- i\beta H - i(\epsilon_1 - \epsilon_2) J_L^3 -i (\epsilon_1 + \epsilon_2) J_R^3    -i (\epsilon_1 +
   \epsilon_2) J_I^3}       \right ) ~,
 \eea
  where $j_L$ and $j_R$ correspond to the spins under the little group $SO(4)$ and $J_I^3$ is a generator of the R-symmetry group $SU(2)$.
   The spectrum of the theory contains the   excitations of elementary fields  and the solitons which correspond to four dimensional instantons.
   The contribution of  elementary excitations of fundamental fields gives rise to perturbative part of the index.
  The perturbative contribution of the vector multiplet to the partition function is given by the following expression
  \bea
  Z_{\mathbb{R}^4 \times S^1}^{pert. vect} (a, \beta, \epsilon_1, \epsilon_2)= \prod\limits_{\alpha \in \Delta} \prod\limits_{k,l=0}^\infty
   \left (1 - e^{i\beta \langle \alpha, a\rangle} e^{ik \beta \epsilon_1} e^{i l\beta \epsilon_2}  \right ) ~,\label{pert-vector-flat}
  \eea
 where $\Delta$ are the   roots and $a=(a_1, a_2, ... , a_r)$ are the asymptotic values (in Cartan) of the scalar field at infinity.
   The function (\ref{pert-vector-flat}) converges if ${\rm Im} (\beta \epsilon_1) >0$ and  ${\rm Im} (\beta \epsilon_2) >0$.
    For a general region  but insisting $\im (\beta \epsilon_1) \neq 0$, $\im(\beta \epsilon_2) \neq 0$, we define $Z_{\mathbb{R}^4 \times S^1}^{pert.vect}$ using a compact notation
    \bea
    Z_{\mathbb{R}^4 \times S^1}^{pert.vect} (a, \beta, \epsilon_1, \epsilon_2) = \prod\limits_{\alpha \in \Delta}  (e^{i \beta \langle \alpha, a\rangle} ; e^{i\beta \epsilon_1},
    e^{i\beta \epsilon_2})_\infty~,\label{def-vector-flat}
    \eea
    which is introduced in \cite{MR2101221}. We collect the conventions and definitions of this special function in Appendix \ref{A-sines}.   The perturbative  contribution of hypermultiplet of mass $m$ is given by
 \bea
 Z_{\mathbb{R}^4 \times S^1}^{pert.hyper} (a, m, \beta, \epsilon_1, \epsilon_2)=   \Big ( \prod\limits_{\mu \in {\cal W}} \prod\limits_{k,l=0}^\infty
   \left (1 - e^{ i \beta (\langle \mu, a\rangle +m)} e^{i(k+1/2)\beta \epsilon_1} e^{i(l+1/2)\beta \epsilon_2}  \right ) \Big)^{-1} ~,\label{pert-hyper-flat}
 \eea
  where ${\cal W}$ are the weights for the representation $\underline{R}$. Again the above expression converges
   if ${\rm Im} (\beta \epsilon_1) >0$ and  ${\rm Im} (\beta \epsilon_2) >0$. In a general region we define it as
   \bea
    Z_{\mathbb{R}^4 \times S^1}^{pert.hyper} (a, m, \beta, \epsilon_1, \epsilon_2) =  (e^{i\beta [ \langle \mu, a\rangle +m +\frac{(\epsilon_1 + \epsilon_2)}{2}]}; e^{i\beta \epsilon_1},
    e^{i\beta \epsilon_2})_\infty^{-1}~,\label{def-pert-hyper-flat}
   \eea
   where we still have to assume that ${\rm Im} (\beta \epsilon_1) \neq 0$ and  ${\rm Im} (\beta \epsilon_2) \neq 0$ as before.

 The spectrum of solitons in the theory on $\mathbb{R}^4 \times S^1$ is given by the instanton partition function. Let us introduce the conventions for instanton
  contributions for vector multiplet and hypermultiplet of mass $m$, as $Z^{inst}_{\mathbb{R}^4 \times S^1}  (a, m, \beta, \epsilon_1, \epsilon_2)$.
   In this work we are not concerned with the explicit form the instanton partition function.
  For example, if we are interested in the $U(N)$ gauge group then
 \bea
  Z^{inst}_{\mathbb{R}^4 \times S^1}(a, m, \beta, \epsilon_1, \epsilon_2) = \sum\limits_{\vec{Y}} q^{|\vec{Y}|} Z_{\vec{Y}} (a, m, \beta, \epsilon_1, \epsilon_2)\label{instanton_PF}
 \eea
 where the sum is over the set of Young diagrams $N$-tuples and $q$ is instanton counting parameter.
   For further details we refer to the literature \cite{Nekrasov:2002qd, Nekrasov:2003rj}.

\section{Partition function on $S^5$}\label{s-sphere}

As a warmup for the discussion of $Y^{p,q}$ spaces we briefly recall the partition function for $S^5$.
 The perturbative part of the partition function for a round $S^5$ has been calculated in \cite{Kallen:2012cs, Kallen:2012va, Kim:2012ava} and
  it has been generalised to the case of squashed $S^5$ in \cite{Lockhart:2012vp, Kim:2012qf, Imamura:2012bm}. The full partition function has been conjectured
   in \cite{Lockhart:2012vp, Kim:2012qf}. Here we run through the logic for the full partition function on $S^5$.

\subsection{Perturbative partition function on $S^5$}

 The perturbative part of the partition function on the squashed $S^5$ for vector multiplet coupled to hypermultiplet
  of mass $m$ in representation $\underline{R}$ is given by the following matrix integral over Cartan subalgebra
  \bea
  \int\limits_{\FR{t}}da~e^{-\frac{8\pi^3 r\varrho}{\gYM^2}\,\Tr[a^2]}
~\frac{{\det}'_{adj}S_3(i a| \omega_1, \omega_2,  \omega_3)}{{\det}_{\underline{R}}S_3(ia + im + \frac{1}{2} (\omega_1 + \omega_2 + \omega_3)| \omega_1, \omega_2, \omega_3)}~,\label{label_S5}
 \eea
  where $\go_1, \go_2, \go_3$ are the squashing parameters for $S^5$ and $\varrho =  \textrm{Vol}_{S^5_{squashed}}/\textrm{Vol}_{S^5}$. The denominator corresponds to the
   contribution of the hypermultiplet, while the numerator that of the vector multiplet.

  It has been pointed out in \cite{Lockhart:2012vp} that using the factorisation properties of $S_3$, the perturbative partition function on $S^5$ can be decomposed into three copies of perturbative answer of the same theory on
    $\mathbb{R}^4 \times S^1$. To see this, we denote the numerator of (\ref{label_S5}) as
 \bea
  Z_{S^5}^{pert.vect} (a, \omega_1, \omega_2, \omega_3) = \prod\limits_{\alpha \in \Delta}
  S_3 (i\langle \alpha, a \rangle | \omega_1, \omega_2, \omega_3)  ~,\label{sk9393}
   \eea
   where we have written ${\det}'_{adj}$ as a product over the roots.
  Using the definition (\ref{def-vector-flat}) and the factorisation property (\ref{app-fact-33}) of the triple sine function, we get
 \bea
&&   Z_{S^5}^{pert.vect} (a, \omega_1, \omega_2, \omega_3) =\big(\prod\limits_{\alpha \in \Delta} e^{-\frac{\pi i}{6} B_{3,3} (i \langle \alpha, a \rangle |\omega_1, \omega_2, \omega_3)}\big)~ Z_{\mathbb{R}^4 \times S^1}^{pert.vect} (ia, \frac{2\pi}{\omega_2}, \omega_1+ \omega_2,  \omega_3) \nn \\
&&\hspace{3.5cm} \times   Z_{\mathbb{R}^4 \times S^1}^{pert.vect} (ia, \frac{2\pi}{\omega_1}, \omega_3+\omega_1,  \omega_2 )
  ~  Z_{\mathbb{R}^4 \times S^1}^{pert.vect} (ia, \frac{2\pi}{\omega_3}, \omega_1+ \omega_3 ,
    \omega_2)~.\label{ident-vectS5}
 \eea
For the hypermultiplet, we denote the denominator of (\ref{label_S5}) as
 \bea
 Z_{S^5}^{pert.hyper} (a, m, \omega_1, \omega_2, \omega_3) =\prod\limits_{\mu \in {\cal W}}
  S^{-1}_3 \Big (i\langle \mu, a \rangle +im + \frac{\omega_1+\omega_2 +\omega_3}{2} | \omega_1, \omega_2, \omega_3 \Big )~,\label{denominator}
 \eea
 where ${\det}_{\underline{R}}$ is written as a product over the weights of $\underline{R}$. We have a similar factorisation
  \bea
 &&Z_{S^5}^{pert.hyper} (a, m, \omega_1, \omega_2, \omega_3) = \nn \\
&&\times \prod\limits_{\mu \in {\cal W}} e^{\frac{\pi i}{6} B_{3,3} (i \langle \mu, a \rangle + im + \frac{\go_1 +\go_2 + \go_3}{2} |\omega_1, \omega_2, \omega_3)}
 ~Z_{\mathbb{R}^4 \times S^1}^{pert.hyper} \Big (ia,
  i m  ,
  \frac{2\pi}{\omega_2}, \omega_1 + \omega_2,  \omega_3 \Big ) \nn \\
&& \times  Z_{\mathbb{R}^4 \times S^1}^{pert.hyper} \Big (ia,
  i m ,
  \frac{2\pi}{\omega_1},  \omega_3+\omega_1,  \omega_2 \Big )
   Z_{\mathbb{R}^4 \times S^1}^{pert.hyper} \Big (ia,
  i m ,
  \frac{2\pi}{\omega_3}, \omega_1 +\omega_3,  \omega_2 \Big )  ~. \label{ident-hyperS5}
  \eea
 In the formulae (\ref{ident-vectS5}), (\ref{ident-hyperS5}) the identification of the parameters $\ep_1,\;\ep_2$ is not unique due to the periodicity
  in the arguments of the flat answer (\ref{pert-vector-flat}), (\ref{pert-hyper-flat}) as well as (\ref{def-vector-flat}) and (\ref{def-pert-hyper-flat}). We have chosen such $\epsilon_1$ and $\epsilon_2$ so that there is
   no shift in the mass of hypermultiplet. A different choice of $\epsilon_1$ and $\epsilon_2$ may imply an additional shift in the mass $m$,  but the combination $im+(\go_1+\go_2+\go_3)/2$ in (\ref{denominator}) is unambiguous.

   \subsection{Full partition function on $S^5$}\label{sec_S5_PF_full}

Using the factorisation properties for perturbative part (\ref{ident-vectS5}) and  (\ref{ident-hyperS5}) we can conjecture
   the structure of the full partition function
    \bea
 && \int\limits_{\FR{t}}da~e^{-\frac{8\pi^3 r\varrho}{\gYM^2}\,\Tr[a^2]}
~Z_{S^5}^{pert.vect} (a, \omega_1, \omega_2, \omega_3) Z_{S^5}^{pert.hyper} (a, m, \omega_1, \omega_2, \omega_3) \nn \\
&& Z_{\mathbb{R}^4 \times S^1}^{inst} \Big (ia,
  i m  ,
  \frac{2\pi}{\omega_2}, \omega_1+ \omega_2,  \omega_3 \Big )
   Z_{\mathbb{R}^4 \times S^1}^{inst} \Big (ia,
  i m ,
  \frac{2\pi}{\omega_1},  \omega_3+ \omega_1,  \omega_2 \Big ) \nn \\
 &&  \times Z_{\mathbb{R}^4 \times S^1}^{inst} \Big (ia,
  i m  ,
  \frac{2\pi}{\omega_3}, \omega_1+\omega_3 ,  \omega_2 \Big) ~.\label{S5_PF_full}
 \eea
   The identification of the equivariant parameters in the instanton part is the same as in the preceding section.
    The instanton partition function on $\mathbb{R}^4 \times S^1$ has the same periodicity in its arguments as the perturbative answer
     and thus there is the same ambiguity in the identification of the equivariant parameters. But the final answer will likewise not be affected by the ambiguities.
   In the next subsection we discuss the potential problems in derivation of the formula (\ref{S5_PF_full}) from the first principles.

  \subsection{Localization calculation}\label{subs-localS}

 Let us briefly sketch the actual localisation calculation which should lead to the above result. Moreover we would like to
  point out some problems in the calculation which are not discussed in the literature.  For the sake of clarity we concentrate only on the vector multiplet, but the generalisation for the hypermultiplet is straightforward. Our review will be rather brief and for further explanations on the setup
    the reader may consult \cite{Kallen:2012va, Qiu:2013pta}.

 For a supersymmetric gauge theory on a simply connected Sasaki-Einstein manifold, the supersymmetry transformation of the vector multiplet can be mapped to the following cohomological complex
  \bea
&&\begin{array}{ll}
  \delta A = i\Psi~, & \gd \Psi = -\iota_{\sreeb} F+ D_A \gs~, \\
  \gd \chi_H^+ = H_H^+~, & \gd H_H^+=-i\L^A_{\sreeb}\chi^+_H-[\gs,\chi_H^+]~, \\
  \gd \gs =-i ~\iota_{\sreeb} \Psi~, &
\end{array}\label{susy_vect_twist}\eea
 Some explanation of the notation is in order. The field $A$ is the connection 1-form, and $F$ its curvature. The remaining fields are: $\Psi$ a fermionic 1-form; $\sigma$ a bosonic scalar; $\chi_H^+$ a fermionic self-dual horizontal 2-form and finally $H_H^+$ a bosonic self-dual horizontal 2-form, all of which are in the adjoint representation. The Sasaki-Einstein structure implies amongst other things that the Reeb vector $\reeb$ is related to the contact 1-form $\gk$ by $\gk=g\reeb$, and that $\reeb$ is a Killing vector. The operator $\iota_{\sreeb}$ stands for the contraction of a differential form with the vector field $\reeb$ and $\L^A_{\sreeb} = L_{\sreeb} + i [~,\iota_{\sreeb} A] $ is the Lie derivative coupled to the gauge connection. The splitting of the differential form into its horizontal and vertical component is done by means of the projectors
     $\kappa \wedge \iota_{\sreeb}$ and $(1- \kappa \wedge \iota_{\sreeb})$, thanks to the property $\iota_\sreeb \kappa =1$ and $\iota_\sreeb d \kappa=0$.
      One can also decompose a horizontal 2-form into its self-dual and anti-self-dual component by using the projectors $(1 \pm \iota_{\sreeb}\star )/2$, where one would find useful the following property $\iota_{\sreeb} \star \omega_p=
      (-1)^p \star (\kappa \wedge \omega_p)$.
       The cohomological complex (\ref{susy_vect_twist}) has been discussed in \cite{Kallen:2012cs} (also see \cite{Nekrasov:1996cz, Baulieu:1997nj} for earlier discussions). The derivation of the mapping of the supersymmetry transformations on to the cohomological complex (\ref{susy_vect_twist}) has
         been given in detail in \cite{Kallen:2012va}.
     Finally the square of the transformations (\ref{susy_vect_twist}) reads
     \bea
  \delta^2 = -iL_{\sreeb} + G_{i(\sigma +\iota_{\sreeb} A )} ~,\label{susy_closure_vect}
 \eea
 where $G_{i(\sigma +\iota_{\sreeb} A )}$ is a gauge transformation with parameter
  $i (\sigma+\iota_{\sreeb} A )$.

Now let us return to the simplest toric Sasaki-Einstein manifold, the round sphere $S^5$, presented as
    \bea
     |z_1|^2 + |z_2|^2 + |z_3|^2 = 1.\nn
    \eea
    There is a 3-torus action $U(1)^3 \times S^5 \rightarrow S^5$ defined
    as $z_i \rightarrow e^{i \alpha_i} z_i,~~i=1,2,3$ and we denote   by $e_a$ the corresponding vector filed.
    For a round $S^5$, the Reeb vector field is
    \bea
     \reeb = e_1 + e_2 + e_3~,
    \eea
   which  corresponds to the $U(1)$-fibre of Hopf fibration $U(1)\to S^5\to \BB{C}P^2$.
 By introducing the following BRST exact term
\bea
 \delta \int\limits_{S^5} {\rm Tr} \left ( 4 \chi_H^+ \wedge \star ( F_H^+ - \frac{1}{2} H_H^+) + \Psi \wedge \star \overline{\delta \Psi} \right )
\eea
 one can show that the partition function of the theory defined by (\ref{susy_vect_twist}) localises to the configuration
   \bea
    \iota_{\sreeb} F=0~,~~~~~F_H^+=0~,~~~~D_A\sigma=0 ~,\label{local-locus}
   \eea
   where in the path integral we choose the contour along imaginary $\sigma$.
  It is natural to generalise $\reeb$ to the case
    \bea
   \reeb = \go_1 e_1 + \go_2 e_2 + \go_3 e_3~,\label{general-Reeb}
  \eea
   where $\omega_i \in \mathbb{R}_{>0}$ for $S^5$. This positivity condition will be replaced by a dual cone condition (\ref{dual_cone_intro}) for general toric Sasaki-Einstein manifolds.
   Of course one now defines the horizontality and self-duality in (\ref{susy_vect_twist}) using the deformed Reeb (\ref{general-Reeb}) and the corresponding contact 1-form.
   After these minor modifications, the theory still localises on the configuration (\ref{local-locus}).

 Since the manifold is simply connected the configuration $A=0$ and $\sigma=\textrm{const}$ is an isolated fixed point and the one loop contribution around it gives rise
    to the result (\ref{sk9393}). The dual cone condition $\omega_i >0$ is important both for the ability to define the contact structure and
     for the regularisation of the triple sine function $S_3$.
    One observes (e.g., see \cite{MR2101221}) that the answer in terms of the triple sine is defined for complex $\omega_i$'s as along as all of them lie on the same side of a line through the origin of $\BB{C}$. Note that the dual cone condition is a just a special case when the corresponding line is the imaginary axis.

     Moreover for the factorisation
     we need ${\rm Im} (\omega_i/ \omega_j) \neq 0$ for $i\neq j$,  see the end of this section for more discussion on this condition.
      If we allow $\reeb=\re\reeb + i \im\reeb$ with $\omega_i \in \mathbb{C}$ then the localisation locus
      becomes
     \bea
    \iota_{\re\sreeb} F=0~,~~~~~F_H^+=0~,~~~~\iota_{\im\sreeb} F - D_A\sigma=0 ~, \label{local-locus-new}
   \eea
  where we use the following notations
  \bea
  \re\reeb = \sum\limits_{i=1}^3 {\rm Re} (\omega_i)\, e_i~,~~~~~~\im\reeb = \sum\limits_{i=1}^3 {\rm Im} (\omega_i)\, e_i~.\label{general-eeio}
  \eea
  We remark that the notion of horizontalness and self-duality of 2-forms are defined with respect to the real part of $\reeb$, which must therefore satisfy the dual cone condition
  ${\rm Re} (\omega_i) >0$.

  For a generic Reeb\footnote{By a generic Reeb, we mean that the ratios ${\rm Re}(\omega_i)/{\rm Re (\omega_j)}$ are irrational for $i \neq j$. In this case
   $\reeb_0$ has only isolated closed orbits, the so called Reeb orbits. These orbits are over the corners of the base polygon.}, there are no closed orbits for $\re \reeb$ except at three points $z_i=1,~i=1,2,3$, and as we saw in (\ref{ident-vectS5}), (\ref{ident-hyperS5}), the perturbative part of the partition function factorises into three factors, each of which is associated with a closed Reeb orbit. The neighbourhood of a closed Reeb orbit say $z_3=1$, can be identified as a solid torus $\BB{C}^2\times S^1$, where $\BB{C}^2$ is parameterised by the inhomogeneous coordinates $z_1/z_3$ and $z_2/z_3$, while $S^1$ has period $\gb=2\pi/\go_3$. The solid torus is \emph{twisted}, i.e. it is presented as $\BB{C}^2\times [0,\gb]\big/\sim$, with the identification being
  \bea
  (\frac{z_1}{z_3},\frac{z_2}{z_3},0)\sim (e^{2\pi i\frac{\go_1-\go_3}{\go_3}}\frac{z_1}{z_3},e^{2\pi i\frac{\go_2-\go_3}{\go_3}}\frac{z_2}{z_3},\gb)~,\nn
  \eea
  which could be equally well written as
  \bea
  (\frac{z_1}{z_3},\frac{z_2}{z_3},0)\sim (e^{2\pi i\frac{\go_1}{\go_3}}\frac{z_1}{z_3},e^{2\pi i\frac{\go_2}{\go_3}}\frac{z_2}{z_3},\gb)~.\label{twisting_solid_torus_S5}
  \eea
  The twisting parameters of the solid tori are thus $\go_{1,2}/\go_3$ and they appear in the Nekrasov partition functions (\ref{def-vector-flat}) and (\ref{def-pert-hyper-flat}).

  The entire $S^5$ may be built by gluing three such solid tori together with appropriately identified twisting parameters, and thus one has a tantalising cutting and gluing construction of the perturbative partition function. As we shall see in the subsequent sections, the same cutting and gluing pattern persists for the $Y^{p,q}$ manifolds. Let us reiterate that complexifying $\go_i$ is a smooth deformation as far as the perturbative part of the partition function is concerned. One needs the conditions $\im \go_i/\go_j\neq0$ only to make each individual factor in (\ref{def-vector-flat}) well-defined, but once all three such factors are in place, one can safely take the $\go_i$ to be real again.
  Hence it would seem reasonable that we should take the idea of complexifying the Reeb vector field seriously and ask what happens to the instanton sector.

  For the factorisation of instanton sector to be true, what one needs to prove is that the only solutions admitted by the set of equations (\ref{local-locus-new})
   are singular ones, more specifically,
  \bea F=D_A\gs=0\nn\eea
  away from the locus of the closed Reeb orbits.
  In other words for generic $\omega_i$'s the equations (\ref{local-locus-new}) do not have smooth solutions but only singular solutions concentrated round the closed Reeb orbits. The contribution of such singular instantons are captured by the corresponding calculation on $\BB{R}^4\times S^1$, and thus it would be again reasonable to glue together three copies of the Nekrasov instanton partition function to obtain the full partition function $S^5$, which is the rationale behind the result in section \ref{sec_S5_PF_full}, as well as the idea pursued in \cite{Lockhart:2012vp} and \cite{Kim:2012ava}.
  Granted this, the instantons are all point like particles that are far apart and propagate along the closed Reeb orbits. In this way, it would suffice to compute the partition function of the instantons on the space $\BB{R}^4\times S^1$. Secondly since $D_A\gs=0$, the moduli of these instantons are given by constant $\gs$, and then one can quote the result of Nekrasov, who computed the instanton partition function of the 5D supersymmetric theory on $\BB{R}^4\times S^1$ \emph{in the Colomb branch} and in the Omega background (in our context, the omega background corresponds to deforming the equations (\ref{local-locus}) to (\ref{local-locus-new}), and eventually to
  the turning on of the equivariant parameters $\ep,\;\ep'$). In the neighbourhood of each closed Reeb orbit, we can identity the equivariant parameters $\ep,\;\ep'$ from the explicit computation of the perturbative partition function.

  However we are unable to prove the absence of smooth solutions for the localisation
   locus (\ref{local-locus-new}).
   We would like to stress that the problem here is not identical to Nekrasov's four dimensional setting. Since (\ref{local-locus-new}) is not an elliptic system of PDEs, and so we are not able to define an appropriate moduli space  so as to perform  a further localisation thereon.

\section{Partition function of $Y^{p,q}$ manifolds}\label{s-Yspace}

In \cite{Qiu:2013pta} we have calculated the perturbative partition function for squashed $Y^{p,q}$ space.
 The main building block for the answer is given in terms of generalised triple sine  function which is
  defined through the regularised infinite product as follows
  \bea
&&S_3^{Y^{p,q}}(x|  \omega_1, \omega_2, \omega_3, \omega_4)=\prod_{(i,j,k,l) \in \Gl^+_{(p,q)}}\Big(i \omega_1+j
\omega_2+k\omega_3+l\omega_4 +x\Big)\nn\\
&&\hspace{3.5cm}  \Big((i+1)\omega_1+(j+1)\omega_2+(k+1)\omega_3+(l+1)\omega_4 -x\Big) ~, \label{triplegensine-def}
\eea
  where the lattice $\Lambda^+_{(p,q)}$ is defined as
  \bea
  \Gl^+_{(p,q)}=\big\{i,j,k,l\in\BB{Z}_{\geq0}\;|\;i(p+q)+j(p-q) -kp- lp=0\big\}~,\label{lattice-1}
\eea
 and $\omega_1, \omega_2, \omega_3, \omega_4$  are equivariant parameters corresponding to
  $U(1)^4$-action on $\mathbb{C}^4$, but due to the lattice constraint above, there are only three effective parameters. For a vector multiplet coupled to a massive hyper with mass $m$ and in representation $\underline{R}$, the perturbative partition function is given by the following matrix integral
\bea
\int\limits_{\FR{t}}da~e^{-\frac{8\pi^3 r\varrho}{\gYM^2}\,\Tr[a^2] }
~\frac{{\det}'_{adj}~S^{Y^{p,q}}_3(i a| \omega_1, \omega_2,  \omega_3, \omega_4)}{{\det}_{\underline{R}}~
S^{Y^{p,q}}_3 (ia + im + \frac{1}{2} (\omega_1 + \omega_2 + \omega_3 + \omega_4)| \omega_1, \omega_2, \omega_3, \omega_4)}~,\label{finale_intro}
\eea
where $\varrho =  \textrm{Vol}_{Y^{p,q}}/\textrm{Vol}_{S^5}$ (with $\textrm{Vol}_{Y^{p,q}}$ being the equivariant volume). For further details and explanations
 we refer the reader to  \cite{Qiu:2013pta}.

For the general case of $L^{a,b,c}$ spaces the lattice (\ref{lattice-1}) becomes
  \bea
  \Gl^+_{(a,b,c)}=\big\{i,j,k,l\in\BB{Z}_{\geq0}\;|\;i a+jb -kc- l (a+b-c)=0\big\}~,\label{lattice-12}
\eea
the corresponding generalised triple sine function $S_3^{L^{a,b,c}}$ gives the perturbative partition function (\ref{finale_intro}) for these spaces.

 \subsection{$T^{1,1}$ case}

 In this subsection  we consider as an illustration the special case of $Y^{p,q}$ manifold
  with $p=1$ and $q=0$. This space is known in the literature as the $T^{1,1}$ space and its metric cone is the conifold. The space $T^{1,1}$ is the quotient of $S^3 \times S^3$ by the diagonal
    $U(1)$, and is in fact the total space of a $U(1)$-bundle over $S^2\times S^2$ with degree 1 and 1.
     The general case of $Y^{p,q}$ space is treated in the appendix \ref{A-gen-sines}.

Even though here one may well specialise the calculation in the appendix \ref{A-gen-sines} for general $p,\,q$,
 we nonetheless go through the steps to give the reader some hands on experience. We have the special function
 \bea S_3^{T^{1,1}} (x | \omega_1, \omega_2, \omega_3, \omega_4)=\prod\limits_{i,j,k,l=0,~i+j=k+l}^\infty
  \big( i \omega_1 + j \omega_2 + k \omega_3 + l \omega_4 + x\big)\big ( x\to \small{\textrm{$\sum\limits_{i=1}^4$}}\go_i-x \big )~,\label{S3_11}\eea
  where the real parts of $\go_i$ are assumed to satisfy the dual cone condition (\ref{app-cone-conditions}), essentially for the ability to use the $\zeta$-function regularisation.

  Let us look at the first factor
  \bea &&\prod\limits_{i,j,k,l=0,~i+j=k+l}^\infty
  ( i \omega_1 + j \omega_2 + k \omega_3 + l \omega_4 + x)\nn\\
  &&\hspace{1cm}=\prod\limits_{i=0}^\infty \prod\limits_{j=0}^\infty \prod\limits_{l=0}^{i+j} (( i \omega_1 + j \omega_2 + (i+j-l)\omega_3 + l \omega_4 + x)\nn\\
   &&\hspace{1cm}=\frac{\prod\limits_{i=0}^\infty \prod\limits_{j=0}^\infty \prod\limits_{l=0}^{\infty} (( i \omega_1 + j \omega_2 + (i+j-l)\omega_3 + l \omega_4 + x)}{\prod\limits_{i=0}^\infty \prod\limits_{j=0}^\infty \prod\limits_{l=i+j+1}^{\infty} (( i \omega_1 + j \omega_2 + (i+j-l)\omega_3 + l \omega_4 + x) }~,\nn\eea
   where the last manipulation is valid if we assume $\re(\go_4-\go_3)>0$, for otherwise the real part of the argument in the bracket will turn negative for large $l$. After shifting the lower limits of the summation indices, we get
  \bea&& \frac{\prod\limits_{i=0}^\infty \prod\limits_{j=0}^\infty \prod\limits_{l=0}^{\infty} (( i \omega_1 + j \omega_2 + (i+j-l)\omega_3 + l \omega_4 + x)}{\prod\limits_{i=0}^\infty \prod\limits_{j=0}^\infty \prod\limits_{l=0}^{\infty} ( i \omega_1 + j \omega_2 + (i+j-i-j -1-l)\omega_3 + (l+ i+ j +1) \omega_4 + x) }\nn\\
  &&=\frac{\prod\limits_{i=0}^\infty \prod\limits_{j=0}^\infty \prod\limits_{l=0}^{\infty} (( i [\omega_1 +\omega_3]+ j [\omega_2+\omega_3] +  l [\omega_4 -\omega_3]+ x)}{\prod\limits_{i=0}^\infty \prod\limits_{j=0}^\infty \prod\limits_{l=0}^{\infty} ( i [\omega_1 +\omega_4]+ j [\omega_2+\omega_4] + l [\omega_4-\omega_3] + \omega_4 - \omega_3 + x) }~.\nn\eea
  Doing the same for the second factor of (\ref{S3_11}), we get
   $$S_3^{T^{1,1}} (x | \omega_1, \omega_2, \omega_3, \omega_4)= \frac{S_3 (x| \omega_1+\omega_3, \omega_2+\omega_3, \omega_4-\omega_3)}{S_3 (x+\omega_4- \omega_3| \omega_1+\omega_4, \omega_2+\omega_4, \omega_4-\omega_3)}~.$$
 After rewriting $S_3^{T^{1,1}}$ as ordinary $S_3$ functions, we can use the factorisation formula (\ref{fact-triplesine}) to get
 \bea
  S_3^{T^{1,1}}(x| \omega_1, \omega_2, \omega_3, \omega_4) = e^{-\frac{\pi i}{6} B_{3,3} (x| \omega_1 + \omega_3, \omega_2 + \omega_3, \omega_4-\omega_3) + \frac{\pi i}{6} B_3(x + \omega_4- \omega_3| \omega_1 + \omega_4, \omega_2 + \omega_4, \omega_4 - \omega_3)} \\
\times  \frac{\prod\limits_{j,k=0}^\infty \left(1 - e^{2\pi i (\frac{x}{\omega_2 + \omega_3} + j \frac{\omega_1 + \omega_3}{\omega_2 + \omega_3} + k \frac{\omega_4 - \omega_3}{\omega_2 + \omega_3})} \right )  \prod\limits_{j,k=0}^\infty \left(1 - e^{2\pi i (\frac{x}{\omega_1 + \omega_3} - (j+1) \frac{\omega_4 - \omega_3}{\omega_1 + \omega_3} - (k+1) \frac{\omega_2 + \omega_3}{\omega_1 + \omega_3})} \right )}
  {\prod\limits_{j,k=0}^\infty \left(1 - e^{2\pi i (\frac{x}{\omega_2 + \omega_4} + j \frac{\omega_1 + \omega_4}{\omega_2 + \omega_4} + (k+1)\frac{\omega_4 - \omega_3}{\omega_2 + \omega_4})} \right )  \prod\limits_{j,k=0}^\infty \left(1 - e^{2\pi i (\frac{x}{\omega_1 + \omega_4} - j \frac{\omega_4 - \omega_3}{\omega_1 + \omega_4} - (k+1) \frac{\omega_2 + \omega_4}{\omega_1 + \omega_4})} \right )}~,\nn\eea
where one should make sure that the imaginary parts of ratios of $\go_i$ be in the correct region for a correct application of (\ref{fact-triplesine}). Instead of spelling out the cumbersome conditions, we rewrite the result as
\bea S_3^{T^{1,1}}(x)&=& e^{-\frac{\pi i}{6} B_{3,3} (x| \omega_1 + \omega_3, \omega_2 + \omega_3, \omega_4-\omega_3) + \frac{\pi i}{6} B_3(x + \omega_4- \omega_3| \omega_1 + \omega_4, \omega_2 + \omega_4, \omega_4 - \omega_3)} \nn\\
  &&\left (e^{2\pi i \frac{x}{\omega_2 + \omega_3}};  e^{2\pi i \frac{\omega_1 + \omega_3}{\omega_2 + \omega_3}}
     e^{2\pi i \frac{\omega_4 - \omega_3}{\omega_2 + \omega_3}} \right )_\infty
  \left ( e^{2\pi i \frac{x}{\omega_1 + \omega_3}};  e^{2\pi i \frac{\omega_4 - \omega_3}{\omega_1 + \omega_3}},
   e^{2\pi i \frac{\omega_2 + \omega_3}{\omega_1 + \omega_3}} \right )_\infty \nn\\
  &&  \left ( e^{2\pi i \frac{x}{\omega_2 + \omega_4}};  e^{2\pi i \frac{\omega_1 + \omega_4}{\omega_2 + \omega_4}},
   e^{2\pi i \frac{\omega_3 - \omega_4}{\omega_2 + \omega_4}} \right )_\infty
  \left ( e^{2\pi i \frac{x}{\omega_1 + \omega_4} } ;    e^{2\pi i \frac{\omega_2 + \omega_4}{\omega_1 + \omega_4}} ,
  e^{2\pi i \frac{\omega_3 - \omega_4}{\omega_1 + \omega_4}} \right )_\infty \nn\eea
  by using a special function defined in (\ref{piecewise_inf}). Now the result is valid as long as the appropriate ratios of $\go_i$ have nonzero imaginary parts.

\subsection{General $Y^{p,q}$ case}

The key piece in (\ref{finale_intro}), namely the generalised triple sine $S^{Y^{p,q}}_3$, has a very interesting factorisation property (\ref{factor_S_3_Ypq}), whose derivation is relegated to the appendix. Using this result we have the factorisation of perturbative answer into four pieces
\bea
&&   Z_{Y^{p,q}}^{pert.vect} (a, \omega_{1,\cdots ,4}) =\big(\prod\limits_{\alpha \in \Delta} e^{B^{Y^{p,q}} (i \langle \alpha, a \rangle |\omega_{1,\cdots,4})}\big)~ \prod^4_{i=1}Z_{\mathbb{R}^4 \times S^1}^{pert.vect} (ia, \gb_i, \ep_i, \ep_i')~, \\
&& Z_{Y^{p,q}}^{pert.hyper} (a, m, \omega_{1,\cdots ,4}) =\nn\\
&&\hspace{1.5cm} \big(\prod\limits_{\mu \in {\cal W}} e^{-B^{Y^{p,q}}(i \langle \mu, a \rangle + im + \frac{\go_1+\cdots+\go_4}{2} |\omega_{1,\cdots, 4})}\big)
 \prod_{i=1}^4 Z_{\mathbb{R}^4 \times S^1}^{pert.hyper} \Big (ia,
  i m,\gb_i,\ep_i,\ep_i'\Big )~,\eea
where $Z_{\mathbb{R}^4 \times S^1}^{pert.vect}$ and $Z_{\mathbb{R}^4 \times S^1}^{pert.hyper}$ are as in (\ref{def-vector-flat}), (\ref{def-pert-hyper-flat}) and the parameters $\gb_i,\ep_i,\ep_i'$ are listed in the four rows of table (\ref{tab_equiv}). The prefactor $B^{Y^{p,q}}$ is a combination of Bernoulli polynomials given in (\ref{Ypq_prefactor_new}).
 It turns out that even this prefactor factorises \cite{Nieri:2013vba}, but we have left this out in the current work since it is not clear how to interpret this factorisation geometrically.

The above formulae fix for us the identification of the parameters $\gb_i,\ep_i,\ep_i'$, which are then used also in the instanton part of the answer. To summarise, our main conjecture is that the pull partition function has the following structure
\bea
&&\int\limits_{\FR{t}}da~e^{-\frac{8\pi^3 r\varrho}{\gYM^2}\,\Tr[a^2] }
 Z_{Y^{p,q}}^{pert.vect} (a, \omega_{1,\cdots ,4})Z_{Y^{p,q}}^{pert.hyper} (a,m, \omega_{1,\cdots ,4})\prod\limits_{i=1}^4 \Big [ Z_{\mathbb{R}^4 \times S^1}^{inst} \left (ia,
  i m  ,
  \beta_i, \epsilon_i,  \epsilon'_i \right ) \Big ]~.\nn\\
  \label{fullans-Ypq}\eea
In the next section we show how the equivariant parameters in table (\ref{tab_equiv}) indeed give us the circumference and the twisting of the solid tori.
We remark that, in the same way as the $S^5$, the identification of the parameters for $Y^{p,q}$ is not unique due to certain periodicity properties.
\bea
  \begin{array}{c|ccc}
      & 2\pi\gb^{-1} & \ep & \ep' \\
      \hline
    z_2=z_4=0 & p\go_1+(p+q)\go_3 & \go_1+\go_2+2\go_3 & \go_4-\go_3 \\
    z_2=z_3=0 & p\go_1+(p+q)\go_4 & \go_1+\go_2+2\go_4 & \go_3-\go_4 \\
    z_1=z_4=0 & p\go_2+(p-q)\go_3 & \go_1+\go_2+2\go_3 & \go_4-\go_3 \\
    z_1=z_3=0 & p\go_2+(p-q)\go_4 & \go_1+\go_2+2\go_4 & \go_3-\go_4 \\
  \end{array}.\label{tab_equiv}\eea

\section{Identification of  parameters on $Y^{p,q}$}\label{s-instantons}

 $Y^{p,q}$ space is toric Sasaki-Einstein manifold.  The whole discussion presented in subsection  \ref{subs-localS} is completely applicable
  for the case of $Y^{p,q}$ space. The localisation of the theory is controlled by the same cohomological complex (\ref{susy_vect_twist}) and by the details of
   the toric contact geometry on $Y^{p,q}$. All the same problem as in $S^5$ case exist for $Y^{p,q}$ space.
   Thus we assume the same setup as subsection \ref{subs-localS}. Below we explain the identification
    of parameters which comes from the geometry and it agrees with the identification coming from the factorisation of special functions (see previous section).

\subsection{Identification of equivariant parameters}

The $Y^{p,q}$ manifold can be presented as a quotient of $S^3\times S^3$ by a free $U(1)_T$ acting on the four coordinates $[z_1,z_2,z_3,z_4]$ with charge $[p+q,p-q,-p,-p]$, where $z_{1,2}$ ($z_{3,4}$) are the coordinates of the first (second) $S^3$.
With this description, we can identify $Y^{p,q}$ as four twisted solid tori glued together, and we shall identify the twisting parameters in this section. We will also give a second derivation of these parameters which is independent of how $Y^{p,q}$ is presented, but only relies on the toric data.

We first find the closed Reeb orbits for a generic Reeb given by $[\go_1,\cdots,\go_4]$ within the dual cone. At a generic point, for the orbit passing it to be closed and of period $\gb$, one must have $ \gb\cdotp\go=\phi[p+q,p-q,-p,-p]+2\pi\vec n$, where $\vec n$ is a set of four integers. Clearly there is no solution to this set of four equations while we only have two variables $\gb,~\phi$. But at points where some (two) of the tori degenerate, we need only satisfy two out of the four equations above and there will be solutions. For example, at $z_2=z_4=0$, we have the equations
\bea
\gb\go_1=\phi(p+q)+k~,~~\gb\go_3=-\phi p+l~,~~~\To~~~
\gb=2\pi\frac{pk+(p+q)l}{p\go_1+(p+q)\go_3}~,\nn
\eea
and by using the coprimeness of $p,~q$, we see that the period of this particular orbit is
\bea
\gb=\frac{2\pi}{p\go_1+(p+q)\go_3}~.\nn
\eea
We would like to map a small neighbourhood of the closed Reeb orbit to the form $\BB{C}^2\times S^1$.
To this end, we need to choose two good coordinates that have zero charge under $U(1)_T=[p+q,p-q,-p,-p]$, we can choose
\bea
 u=z_1z_2z_3^2~,~~~v=\frac{z_4}{z_3}~,\nn
 \eea
and the fixed point correspond to $u=v=0$. Since $z_1,~z_3\neq0$, this is a good change of coordinates. Then for every $\gt$ degree one travels along the Reeb vector field, the transverse coordinates $u,~v$ rotate according to
\bea
u\to e^{i\gt(\go_1+\go_2+2\go_3)}u~,~~~v\to e^{i\gt (\go_4-\go_3)}v~,\nn
\eea
so the solid tori is twisted as
\bea
 (u,v,0)\sim (e^{i\gb\ep}u,e^{i\gb\ep'}v,\gb)~,~~ \textrm{where}~~\ep=\go_1+\go_2+2\go_3~,~~\ep'=\go_4-\go_3~.\nn
 \eea
The analysis at the other three loci of closed Reeb orbits is entirely similar and the result is collected in the table (\ref{tab_equiv}).
These parameters are precisely the combination of equivariant parameters appearing in each factor of the factorisation (\ref{factor_S_3_Ypq}) of the special
 function $S_3^{Y^{p,q}}$.

Next, we would like to start from the toric data, and reach the same identification of parameters, with the goal that one should be able to read off the partition function on a toric SE manifold solely from the toric diagram.

\subsection{Identification of equivariant parameters-from toric diagram}\label{sec_Ioepftd}

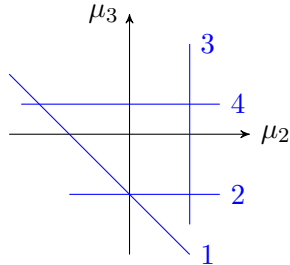
\begin{figure}[h]
\begin{center}
\begin{tikzpicture}[scale=.8]
\draw [->] (-2,0) -- (2,0) node [right] {\small$\mu_2$};
\draw [->] (0,-2) -- (0,2) node [left] {\small$\mu_3$};

\draw [-,blue] (-2,1) -- (1,-2) node [right] {\small$1$};
\draw [-,blue] (-1,-1) -- (1.5,-1) node [right] {\small$2$};
\draw [-,blue] (1,-1.5) -- (1,1.5) node [right] {\small$3$};
\draw [-,blue] (-1.8,0.5) -- (1.5,0.5) node [right] {\small$4$};

\end{tikzpicture}\caption{The polytope cone for $p=3$, $q=2$, projected onto the plane $\mu_1=1+\mu_2+\mu_3$, the cone is the area enclosed by the four lines.}\label{fig_toric_cone}
\end{center}
\end{figure}

As it was briefly reviewed in the introduction, the $Y^{p,q}$ space is toric Sasaki-Einstein. Its metric cone (K\"ahler) can be thought of as a $T^3$-fibration over a polytope cone, defined by four inequalities
\bea
&&\hspace{3cm} \vec \mu\cdot \vec v_i\geq0~,~~~i=1\cdots 4~,~~~\vec{\mu}=[\mu_1,\mu_2,\mu_3]~,\nn\\
&&\vec v_1=[1,0,0]~,~~\vec v_2=[1,-1,0]~,~~\vec v_3=[1,-2,-p+q]~,~~\vec v_4=[1,-1,-p]~,\label{four_normal}
\eea
the $v_i$'s are the inward pointing normal of the four faces of the cone. In Figure \ref{fig_toric_cone}, we draw the projection of the cone onto a plane $\mu_1=1+\mu_2+\mu_3$.
Note that by choosing a different set of three transitively acting $U(1)$'s, one also obtains a different set of $v_i$, here we follow the choice from \cite{Martelli:2004wu}.

The following plane
\bea
\vec{\reeb}\cdot\vec{\mu}=\frac12~\label{cone_base_eqn}
\eea
cuts the polytope cone at a polygon provided the dual cone condition (\ref{dual_cone_intro}) is satisfied.
The $Y^{p,q}$ space is a $T^3$ fibration over this polygon, in particular, at an edge $i$ the torus corresponding to $v_i$ degenerates, but the degeneration is such  that
 whole space is a smooth manifold. More concretely, a polytope cone is said to be $\emph{good}$ (see equation (2.4) in \cite{2001math......7201L}) if at each intersection of two faces, say $i$ and $i+1$, with normals $v_i$, $v_{i+1}$, there exists a third integer entry vector $u$, such that
\bea
 [v_i,v_{i+1},u]\in SL_3(\BB{Z})~.\nn
\eea
It is further shown in Theorem 2.18 of \cite{2001math......7201L} that the 5D compact connected toric contact manifolds with non-free torus action are in 1-1 correspondence with
good moment map cones.

It is easy to check that the four normals in (\ref{four_normal}) define a good cone, thus $Y^{p,q}$ space is a smooth torus fibration over the base polygon like the one in Figure \ref{fig_toric_cone}.

At a vertex, say at the intersection of edge 3 and 4, two out of three tori degenerates, thus the neighbourhood of such a vertex is diffeomorphic to a twisted solid torus $\BB{C}^2\times [0,\gb]\big/\sim$, where $\{0\}\times[0,\gb]/\sim$ is a closed Reeb orbit and it corresponds to the vertex, while the two tori that degenerate act as rotations of the two $\BB{C}$'s.

To obtain the period $\gb$ of the non-vanishing torus, the equation one needs to solve is
\bea
\gb\vec{\reeb}=2\pi\vec n+\gb\ep\vec{v}_3+\gb\ep'\vec{v}_4~,\nn
\eea
where $\vec n$ is integer valued 3-vector satisfying $\det[\vec n,\vec v_3,\vec v_4]=1$ (guaranteed by the goodness of the cone), while $\ep,\ep'$ are the equivariant parameters. The reasoning of this equation is the following, at this vertex, we decompose $\reeb$ in the basis $\vec n$ and $\vec v_{3,4}$, i.e. into three parts, two of which degenerate and a third remains non-degenerate.
The solution is clearly
\bea [\gb,\gb\ep,\gb\ep']^T=2\pi\big[\vec{\reeb}~,-\vec{v}_3,-\vec{v}_4\big]^{-1}\vec{n}~,\label{solving_period}\eea
which is easily solved to give
\bea
&&\frac{\gb}{2\pi}=\frac{\det[\vec n,\vec v_3,\vec v_4]}{\det[\vec\reeb,\vec v_3,\vec v_4]}=\frac{n_1(p+q)+qn_2+n_3}{(p+q)\reeb_1+q\reeb_2+\reeb_3}~,\nn\\
&&\frac{\gb{\ep}_1}{2\pi}=\frac{\det[\vec n,\vec \reeb,\vec v_4]}{\det[\vec\reeb,\vec v_3,\vec v_4]}=\frac{(-p\reeb_2+\reeb_3)n_1+(p\reeb_1+\reeb_3)n_2-(\reeb_1+\reeb_2)n_3}{(p+q)\reeb_1+q\reeb_2+\reeb_3}~,\nn\\
&&\frac{\gb\ep_2}{2\pi}=\frac{\det[\vec n,\vec v_3,\vec \reeb]}{\det[\vec\reeb,\vec v_3,\vec v_4]}=\frac{((p-q)\reeb_2-2\reeb_3)n_1+((q-p)\reeb_1-\reeb_3)n_2+(2\reeb_1+\reeb_2)n_3}{(p+q)\reeb_1+q\reeb_2+\reeb_3}~.\nn\eea
We can simply choose $\vec n=[0,0,1]$, so the period $\gb$ is
\bea \frac{\gb}{2\pi}=\frac{1}{(p+q)\reeb_1+q\reeb_2+\reeb_3}~,\nn\eea
and also
\bea \ep=-\reeb_1-\reeb_2~,~~~\ep'=2\reeb_1+\reeb_2~.\nn\eea
If we choose a different $\vec n$, then $\ep,\ep'$ will be shifted by integer multiples of $2\pi\gb^{-1}$.

Finally, we still need to figure out the mass shift for the hyper-multiplet. This shift was derived in appendix B of  \cite{Qiu:2013pta}, which only works with the Sasaki-Einstein metric, but is applicable for any Killing vector field $X$. With the Sasaki-Einstein metric one can build a convenient spin representation in terms of horizontal differential forms by using the horizonal K\"ahler Einstein structure. With this representation $L_X^s$ is given by the usual Lie derivative plus a shift $L_X^s=L_X+if_X$. We assume either $X$ has a zero or can be decomposed into a linear combination commuting Killing vectors, each of which has a zero. Take one zero of $X$, one can assume that close to this zero $X$ is a linear combination of rotations of $\BB{C}$'s, then for each such rotation of degree 1, one gets a factor of $1/2$ for $f_X$.

To illustrate this, we decompose $\reeb$ into $\reeb=\gl_1\vec v_1+\gl_2\vec v_2+\gl_3\vec v_3$ (one can choose any three $\vec v_i$). The vector field corresponding to $\vec v_1$ vanishes exactly at face 1, and of degree 1, thus one gets from it a shift of $\gl_1/2$. Thus the shift in mass in the formula (\ref{finale_intro}) (or its analog for $L^{a,b,c}$ spaces)
 is written as follows
\bea
 im+\frac{1}{2}\sum_{i=1}^3\gl_i=im+\frac{1}{2}[1,1,1][\vec v_1,\vec v_2,\vec v_3]^{-1}\vec\reeb~.\nn
 \eea
From our particular choice of $\vec v_i$, we get the shifted mass $im+1/2\reeb_1$.

As a check, by reexpressing $\reeb_{1,2,3}$ in terms of the parameters $\go_{1,\cdots,4}$ (for the derivation of the following relation, see subsection 3.4 of \cite{Qiu:2013pta}, our $\reeb_i$ is denoted as $b_i$ there)
\bea \reeb_1=\sum\limits_{i=1}^4\go_i~,~~~\reeb_2=-\go_1-\go_2-2\go_4~,~~~\reeb_3=-p\go_2+(q-p)\go_4~,\nn\eea
we get
\bea
&&\frac{\gb}{2\pi}=\frac{1}{p\go_1+(p+q)\go_3}~,~~~\ep=\go_4-\go_3~,~~~\ep'=\go_1+\go_2+2\go_3~,\nn\eea
One sees that the mass shift is $im+1/2\sum_{i=1}^4\go_i~$, exactly the one obtained from the explicit localisation calculation.
Furthermore, we here have chosen the corner that corresponds to the locus $z_2=z_4=0$ as in the previous section. One can do a similar exercise for other corners of the toric diagram and get the table (\ref{tab_equiv}).  Away from the Sasaki-Einstein metric, extra background fields have to be turned on in order to maintain the supersymmetry, but it seems that the shift we obtained above continues to hold.

Provided that we have chosen the concrete $\epsilon_i$ and $\epsilon'_i$ the shift in the mass for the flat contributions can be easily calculated using the formula (\ref{shift_toric}).
 \emph{It was with this explicit choice} of $\ep,\ep'$ that we wrote the partition function in (\ref{S5_PF_full}) and (\ref{fullans-Ypq}).
Making a different choice of $\vec n$ in (\ref{solving_period}), one will get instead
\bea
 \frac12(\ep+\ep')~\rightarrow~\frac12(\ep+\ep')+\frac{k}{2}\frac{2\pi}{\gb}~,~~~k\in\BB{Z}~.\nn
\eea

\section{Summary}\label{s-summary}

In this work we continue to study the partition function for 5D supersymmetric Yang-Mills theory on the toric Sasaki-Einstein manifold
 $Y^{p,q}$. We show that the perturbative answer factorises in four pieces corresponding to the perturbative answer on $\mathbb{R}^4 \times S^1$.
  This factorisation allows us to identify the equivariant parameters and conjecture also the instanton part of the result, which consists of four copies of Nekrasov instanton partition function on $\mathbb{R}^4 \times S^1$. We also provided the derivation of the identification of parameters from the toric data.
  Our conjectured result relies on the absence of smooth instantons, which is quite widely accepted in the literature, however
   at the moment we cannot prove this fact from the first principles.
    Given this, we obtained the exact partition function for supersymmetric theory for an infinite class of spaces, and it would be interesting to study
  further the properties of the partition function, e.g. such as degenerations, factorisations  and possible modularity.

  It is natural to assume that the result (\ref{fullans-Ypq_intro}) holds for more general toric Sasaki-Einstein manifolds $L^{a,b,c}$ where
   the equivariant parameters can also be deduced from the appropriate toric diagram. However   the proof of the factorisation of $S_3^{L^{a,b,c}}$ turns out a bit tricky and is still under investigation, therefore we shall refrain from further speculations for now.

  It would be interesting to understand  our results from  the point of view of superconformal index of 6 dimensional $(2,0)$ theory on
   $Y^{p,q} \times S^1$ (or nontrivial $S^1$ vibration over $Y^{p,q}$).  Also the relation of our result to the refined topological string theory
    as in \cite{Lockhart:2012vp} remains to be investigated. Many observations for $S^5$ from \cite{Lockhart:2012vp} have straightforward
     generalisation for $Y^{p,q}$ spaces.

\bigskip
{\bf Acknowledgements} We thank   Tobias Ekholm, Guglielmo Lockhart,  Sara Pasquetti, Vasily Pestun,
Cumrun Vafa and Edward Witten for discussions.
 We thank Fabrizio Nieri, Sara Pasquetti, Filippo Passerini and Alessandro Torrielli for sharing their work \cite{Nieri:2013vba} prior to its publication.
We thank the Galileo Galilei Institute for Theoretical Physics, Florence
 for hospitality and partial support during the initial stage of this project.
 M.Z. thanks  the Simons Center for Geometry and Physics for hospitality and  the 11th  Simons workshop in mathematics and physics for providing opportunities for numerous fruitful discussions.
The research of J.Q. is supported by the Luxembourg FNR grant PDR 2011-2, and by the UL grant GeoAlgPhys 2011-2013. The research of M.Z. is supported in part by Vetenskapsr\r{a}det under grant $\sharp$ 2011-5079.

\appendix
\section{Properties of double and triple sines}\label{A-sines}

In this appendix we summarise the relevant properties of multiple sine and multiple gamma functions.  For the detailed exposition of the subject see
 \cite{MR2010282, MR2101221}.

 The multiple zeta function is defined as
 \bea
  \zeta_r (z, s| \omega_1, ... , \omega_r) =\sum\limits_{n_1, .... , n_r =0}^{\infty} (n_1 \omega_1 + ... + n_r \omega_r + z)^{-s} \label{app-zeta}
 \eea
for $z \in \mathbb{C}$, ${\rm Re} ~s > r$ and we assume that ${\rm Re} ~\go_i > 0$. By analytic continuation $\zeta_r$ can be extended to a meromorphic
  function of $s\in\mathbb{C}$.
 The multiple gamma function is defined by the following expression
 \bea
  \Gamma_r (z | \omega_1, ... , \omega_r) = \exp \left ( \frac{\partial}{\partial s} \zeta_r (z, s| \omega_1, ... , \omega_r) |_{s=0} \right )~.
 \eea
  This implies that we can represent $\Gamma_r$ as a regularised infinite product
 \bea
   \Gamma_r (z | \omega_1, ... , \omega_r)  = \prod\limits_{n_1, ... , n_r =0}^{\infty}  (n_1 \omega_1 + ... + n_r \omega_r + z)^{-1}~.\label{A-gamma-prod}
 \eea
Throughout the paper we always consider the regularised infinite products even if it is not stated explicitly. The multiple sine is  defined as follows
 \bea
  S_r (z| \omega_1, ... , \omega_r) =   \Gamma_r (z | \omega_1, ... , \omega_r)^{-1}      \Gamma_r (\omega_1 + ... + \omega_r - z | \omega_1, ... , \omega_r)^{(-1)^r} ~.\label{app-def-sine}
 \eea
  Using (\ref{A-gamma-prod}) we arrive at the following infinite product representation of $S_r$
  \bea
    &&S_r (z| \omega_1, ... , \omega_r) =  \nn \\
   && \prod\limits_{n_1, ... , n_r =0}^{\infty} ((n_1+1) \omega_1 + ... + (n_r+1) \omega_r    - z)
      (n_1 \omega_1 + ... + n_r \omega_r + z)^{(-1)^{r+1}}~. \label{app-triplesine}
  \eea
  We are interested in $\Gamma_r$ and $S_r$ in the case when $r=2$ and $r=3$. Using the above definitions it is straightforward to
   derive formulae like
    \bea
 \Gamma_3 (z+ \omega_2| \omega_1, \omega_2, \omega_3) = \frac{1}{\Gamma_2(z|\omega_1, \omega_3)} \Gamma_3(z| \omega_1, \omega_2, \omega_3)\label{app-relat-1}
\eea
and
 \bea
  S_3(z|\omega_1, \omega_2, \omega_3) = \frac{\Gamma_2(\omega_1 + \omega_3 - z| \omega_1, \omega_3)}{ \Gamma_3(z|\omega_1, \omega_2, \omega_3) \Gamma_3(\omega_1 + \omega_3 -z |\omega_1, \omega_2, \omega_3)}~.\label{app-relat-2}
 \eea
The functions
 $S_2$ and $S_3$ admit the following important factorization \cite{MR2101221}. If  ${\rm Im} (\omega_1/\omega_2) > 0$ then $S_2$ can be factorized as
\bea
S_2(z|\go_1,\go_2) = e^{\frac{\pi i}{2}B_{2,2}(z|\go_1,\go_2)}\frac{\prod\limits_{j=0}^{\infty}\big(1- e^{2\pi i(z/\go_2+j \go_1/\go_2)}\big)}{\prod\limits_{j=0}^{\infty}\big(1-e^{2\pi i(z/\go_1-(j+1)\go_2/\go_1)}\big)}~,\label{fact-doublesine}
\eea
 and similar relation can be written for the region ${\rm Im} (\omega_2/\omega_1) > 0$. For $S_3$
if ${\rm Im} (\omega_1/\omega_2) > 0$,  ${\rm Im} (\omega_1/\omega_3) > 0$ and ${\rm Im} (\omega_3/\omega_2) > 0$ then we have the following factorisation
  \bea
 &&S_3 (z | \omega_1, \omega_2, \omega_3) = e^{-\frac{\pi i}{6} B_{3,3} (z|\omega_1, \omega_2, \omega_3)} \nn\\
&&\times  \frac{\prod\limits_{j,k=0}^\infty (1- e^{2\pi i (z/\omega_2 + j \omega_1/\omega_2 + k \omega_3/
  \omega_2)}) \prod\limits_{j,k=0}^\infty (1 - e^{2\pi i (z/\omega_1 - (j+1) \omega_3/\omega_1 - (k+1) \omega_2/\omega_1)}) }{\prod\limits_{j,k=0}^\infty (1- e^{2\pi i (z/\omega_3 + j \omega_1/\omega_3 -
   (k+1) \omega_2/\omega_3)})}~, \label{fact-triplesine}
    \eea
    and similar expressions can be obtained for other regions. In (\ref{fact-doublesine}) and (\ref{fact-triplesine}) $B_{2,2}$ and $B_{3,3}$ are the
     Bernoulli polynomials
     and they defined as follows
    \bea
   &&  B_{2,2} (z| \omega_1, \omega_2) = \frac{z^2}{\omega_1 \omega_2} - \frac{\omega_1+\omega_2}{\omega_1 \omega_2} z + \frac{\omega_1^2 + \omega_2^2 +
      3 \omega_1 \omega_2}{6 \omega_1 \omega_2}~,\\
     && B_{3,3}(z| \go_1, \go_2, \go_3) = \frac{z^3}{\go_1 \go_2 \go_3} -\frac{3(\go_1 + \go_2 + \go_3)}{2\go_1 \go_2 \go_3} z^2 \\
     &&+
      \frac{\go_1^2 + \go_2^2 + \go_3^2 + 3(\go_1 \go_2 + \go_1 \go_3 + \go_2 \go_3)}{2 \go_1 \go_2 \go_3}z
     - \frac{(\go_1 + \go_2 + \go_3)(\go_1 \go_2 + \go_1 \go_3 + \go_2 \go_3)}{4\go_1 \go_2 \go_3}~. \nn
    \eea
  Following \cite{MR2101221} we define the following meromorphic function of $z$
 \bea
 &&(z; q_1, ... , q_r)_{\infty} \nn\\
 &&=  \left \{ \begin{array}{l}
 \prod\limits_{n_1, ..., n_r =0}^{\infty} (1 - z q_1^{n_1} q_2^{n_2} ... q_r^{n_r})~,~~~|q_1|<1~,|q_2|<1~,~...~, |q_r|<1 \\
  \prod\limits_{n_1, ..., n_r =0}^{\infty} (1 - z q_1^{-n_1-1} q_2^{n_2} ... q_r^{n_r})^{-1}~,~~~|q_1|>1~,|q_2|<1~,~...~, |q_r|<1 \\
 ~~~~~~~~~~~~~  \cdots \\
    \prod\limits_{n_1, ..., n_r =0}^{\infty} (1 - z q_1^{-n_1-1} q_2^{-n_2-1} ... q_r^{-n_r-1})^{(-1)^r}~,~~~|q_1|>1~,|q_2|>1~,~...~, |q_r|>1
  \end{array} \right .\label{piecewise_inf}
 \eea
  This function is not defined if any one of the $q$'s has norm 1.  Moreover the function is clearly invariant under the permutation of $q$'s, and it
   satisfies the following functional equations
   \bea
 (z; q_1, ... , q_r)_{\infty} = \frac{1}{ (q_{j}^{-1}z; q_1, ..., q_j^{-1}, ... , q_r)_{\infty} }~,\label{app-relations-1}\\
 (q_j z; q_1, ... , q_r)_{\infty} = \frac{(z; q_1, ... , q_r)_{\infty} }{(z; q_1, ... ,q_{j-1}, q_{j+1}, ...,  q_r)_{\infty} }~.\label{app-relations-2}
   \eea
  Using these functions we can rewrite the factorisation (\ref{fact-doublesine})   as
  \bea
   S_2(z|\go_1,\go_2) = e^{\frac{\pi i}{2}B_{2,2}(z|\go_1,\go_2)} (e^{2\pi i z/\go_2}; e^{2\pi i \go_1/\go_2})_{\infty}  (e^{2\pi i z/\go_1}; e^{2\pi i \go_2/\go_1})_{\infty}~,\label{app-fact-22}
  \eea
   which is valid as long as ${\rm Im}(\go_1/\go_2)\neq 0$. Analogously we rewrite the factorisation  (\ref{fact-triplesine}) as
   \bea
 &&  S_3 (z | \omega_1, \omega_2, \omega_3) = e^{-\frac{\pi i}{6} B_{3,3} (x|\omega_1, \omega_2, \omega_3)}
   (e^{2\pi i z/\go_2} ; e^{2\pi i \go_1/\go_2}, e^{2\pi i \go_3/\go_2})_\infty \nn \\
&&\times (e^{2\pi i z/\go_1} ; e^{2\pi i \go_3/\go_1}, e^{2\pi i \go_2/\go_1})_\infty
   (e^{2\pi i z/\go_3} ; e^{2\pi i \go_1/\go_3}, e^{2\pi i \go_2/\go_3})_\infty~,\label{app-fact-33}
   \eea
 which is valid  if ${\rm Im} (\omega_1/\omega_2) \neq 0$,  ${\rm Im} (\omega_1/\omega_3) \neq 0$ and ${\rm Im} (\omega_3/\omega_2) \neq 0$.

 \section{Properties of generalized triple sines}\label{A-gen-sines}

 If one looks at the expressions (\ref{app-zeta})-(\ref{app-triplesine}) then it is natural to generalise these functions to the case when the sums and
  products are taken over more general lattices. But we are unaware of any systematic study of such generalisations, thus we concentrate in this appendix
   on the specific lattice and we derive some original relations for the corresponding special functions.

 Let us define the generalised triple zeta function associated to $Y^{p,q}$ manifold as
 \bea
  \zeta^{Y^{p,q}}_3 (z, s| \omega_1, ... , \omega_4) =\sum\limits_{i, j, k, l \in  \Gl^+_{(p,q)}} (i \omega_1 + j \omega_2 +k \omega_3  + l \omega_4+ z)^{-s}~, \label{app-general-zeta}
 \eea
  where the lattice is defined as
    \bea
  \Gl^+_{(p,q)}=\big\{i,j,k,l\in\BB{Z}_{\geq0}\;|\;i(p+q)+j(p-q) -kp- lp=0\big\}~,\label{app-lattice-1}
\eea
 for two coprime integers $p>q$. Due to the lattice condition we effectively sum over a triplet of integers and this is why we refer to this function as the
  generalised triple zeta function. In the above expression
  $\zeta^{Y^{p,q}}_3$ is defined for  $z \in \mathbb{C}$, ${\rm Re} ~s > 3$, and by analytic continuation, it can be extended to $s \in \mathbb{C}$. 
  The conditions for $\omega_i$ are discussed below shortly.
   
 The corresponding generalised triple gamma function is defined by the following expression
 \bea
  \Gamma_3^{Y^{p,q}} (z | \omega_1, ... , \omega_4) = \exp \left ( \frac{\partial}{\partial s} \zeta_3^{Y^{p,q}} (z, s| \omega_1, ... , \omega_4) |_{s=0} \right )~.\label{ap-defgengamma}
 \eea
  The generalised triple sine is defined as follows
 \bea
&&  S_3^{Y^{p,q}}(z| \omega_1, \omega_2, \omega_3, \omega_4) \nn\\
  &&\hspace{1cm}=  [\Gamma^{Y^{p,q}}_3( z|\omega_1, \omega_2, \omega_3, \omega_4)  \Gamma^{Y^{p,q}}_3( \omega_1 + \omega_2 + \omega_3 + \omega_4-z|\omega_1, \omega_2, \omega_3, \omega_4)]^{-1}.\label{app-def-genersine}
 \eea
The function $S_3^{Y^{p,q}}$ can be written as a regularised infinite product (\ref{triplegensine-def}).

Let us analyse the lattice $\Gl^+_{(p,q)}$.  Since $p$ and $q$ are coprime and $p> q$, the lattice condition (\ref{app-lattice-1})
  can be solved as follows
 $$ m\geq 0~,~i \geq0~: ~j= i + mp~,~k+l = 2i + m (p-q) ~,  $$
 $$ m \geq 0~,~j \geq 0~:~i = j+ (m+1)p~,~k+l = 2j  + (m+1) (p+q)~.$$
 Therefore the sum (\ref{app-general-zeta}) can be rewritten as follows
 $$\sum\limits_{i,j,k,l \in \Lambda^+} (i \omega_1 + j \omega_2 + k \omega_3 + l \omega_4 +x)^{-s}=$$
$$= \sum\limits_{i=0}^\infty \sum\limits_{m=0}^\infty \sum\limits_{l=0}^{2i + m (p-q)}
  \Big ( i (\omega_1 + \omega_2 + 2 \omega_3) + m (p\omega_2 + (p-q) \omega_3) + l (\omega_4-\omega_3) +x \Big )^{-s}  $$
  $$+  \sum\limits_{j=0}^\infty \sum\limits_{m=0}^\infty \sum\limits_{l=0}^{2j + (m+1) (p+q)}
  \Big ( j (\omega_1 + \omega_2 + 2\omega_3) + m (p \omega_1 + (p+q)\omega_3) + l (\omega_4 - \omega_3) + x + p\omega_1 + (p+q) \omega_3 \Big )^{-s}~.$$
   This leads to the following relation among zeta functions
   \bea
  &&   \zeta^{Y^{p,q}}_3 (z, s| \omega_1, ... , \omega_1) =
     \zeta_3(z, s | \omega_1 +\omega_2 + 2\omega_3, p\omega_2 + (p-q)\omega_3, \omega_4 - \omega_3)  \nn \\
    && - \zeta_3 (z + \omega_4 - \omega_3, s| \omega_1 +\omega_2 + 2\omega_4, p\omega_2 + (p-q)\omega_4, \omega_4 - \omega_3) \nn  \\
&&+ \zeta_3(z + p\omega_1 + (p+q)\omega_3, s | \omega_1 + \omega_2 + 2 \omega_3, p\omega_1 +(p+q) \omega_3, \omega_4 - \omega_3) \label{app-rel-zeta} \\
&& -\zeta_3 (z + \omega_4 - \omega_3 + p\omega_1 + (p+q)\omega_4, s| \omega_1 +\omega_2 + 2\omega_4, p\omega_1 + (p+q)\omega_4, \omega_4 - \omega_3) \nn
   \eea
    and by analytical continuation the relation will hold outside of the region ${\rm Re} ~s > 3$.  We need to assume that
    \bea
     {\rm Re} (p\omega_2 + (p-q)\omega_{3,4}) >0~,~~~~
     {\rm Re} ( p\omega_1 +(p+q) \omega_{3,4})>0~, \label{app-cone-conditions}
      \eea
    and
    \bea
     {\rm Re} (  \omega_1 + \omega_2 + 2 \omega_{3,4} ) >0~,~~~~
    {\rm Re} (\go_4-\go_3) >0~,\label{app-zeta-cond}
    \eea
     for (\ref{app-rel-zeta}) to be well-defined.  Indeed the conditions (\ref{app-cone-conditions}) are equivalent to the dual cone conditions (\ref{dual_cone_intro}), which is discussed in \cite{Qiu:2013pta} (see
      the equation (42) and the discussion around) and it is essential for the partition function being well-defined.
      The first condition in (\ref{app-zeta-cond}) is a direct consequence of the two in (\ref{app-cone-conditions}). However the last one in (\ref{app-zeta-cond})
      has to be imposed additionally for the factorisation to work.
    Using the definition (\ref{ap-defgengamma}) and the  relation (\ref{app-rel-zeta})  between zeta functions we arrive at  the following relation between gamma functions
    \bea
 &&  \Gamma^{Y^{p,q}}_3( z|\omega_1, \omega_2, \omega_3, \omega_4) = \nn \\
  && \frac{\Gamma_3(z| \omega_1 +\omega_2 + 2\omega_3, p\omega_2 + (p-q)\omega_3, \omega_4 - \omega_3)}{\Gamma_3 (z + \omega_4 - \omega_3| \omega_1 +\omega_2 + 2\omega_4, p\omega_2 + (p-q)\omega_4, \omega_4 - \omega_3)}  \label{app-gamma-rel}\\
&& \times \frac{ \Gamma_3(z + p\omega_1 + (p+q)\omega_3 | \omega_1 + \omega_2 + 2 \omega_3, p\omega_1 +(p+q) \omega_3, \omega_4 - \omega_3)}{ \Gamma_3 (z + \omega_4 - \omega_3 + p\omega_1 + (p+q)\omega_4| \omega_1 +\omega_2 + 2\omega_4, p\omega_1 + (p+q)\omega_4, \omega_4 - \omega_3)}~.\nn
\eea
   Using the relation (\ref{app-gamma-rel}) and the definitions (\ref{app-def-sine}), (\ref{app-def-genersine})
  \bea
  S_3^{Y^{p,q}}(z| \omega_1, \omega_2, \omega_3, \omega_4)
&=&\frac{S_3(z| \omega_1 +\omega_2 + 2\omega_3, p\omega_2 + (p-q)\omega_3, \omega_4 - \omega_3)}{S_3(z + \omega_4 - \omega_3| \omega_1 +\omega_2 + 2\omega_4, p\omega_2 + (p-q)\omega_4, \omega_4 - \omega_3)}\nn\\
&&~~\times\frac{S_3(z| \omega_1 + \omega_2 + 2 \omega_3, p\omega_1 +(p+q) \omega_3, \omega_4 - \omega_3)}{
S_3(z + \omega_4 - \omega_3| \omega_1 +\omega_2 + 2\omega_4, p\omega_1 + (p+q)\omega_4, \omega_4 - \omega_3)}\nn\\
&&\times\frac{S_2(z+\go_4-\go_3| \omega_1 +\omega_2 + 2\omega_4,\omega_4 - \omega_3)}{S_2(z| \omega_1 +\omega_2 + 2\omega_3, \omega_4 - \omega_3)}~,\label{app-sine-sine}
\eea
where we have left out some steps during the derivation, involving the use of the relations like (\ref{app-relat-1}) and (\ref{app-relat-2}).

 Using the factorisation (\ref{app-fact-22}) and (\ref{app-fact-33}) we can rewrite (\ref{app-sine-sine}) as follows
 \bea
 && S^{Y^{p,q}}_3 (z| \omega_1, \omega_2, \omega_3, \omega_4) = e^{B^{Y^{p,q}} (z|\go_1, \go_2, \go_3, \go_4)} \nn\\
&& \times (e^{\frac{2\pi i z}{p \go_1 + (p+q) \go_3}}; e^{\frac{2\pi i (\go_1 + \go_2 + 2 \go_3)}{p \go_1 + (p+q) \go_3}}, e^{\frac{2\pi i(\go_4 - \go_3)}{p \go_1 + (p+q) \go_3}})_\infty
  (e^{\frac{2\pi i z}{p \go_1 + (p+q) \go_4}}; e^{\frac{2\pi i (\go_1 + \go_2 + 2 \go_4)}{p \go_1 + (p+q) \go_4}}, e^{\frac{2\pi i(\go_3 - \go_4)}{p \go_1 + (p+q) \go_4}})_\infty \nn \\
&&\times  (e^{\frac{2\pi i z}{p \go_2 + (p-q) \go_4}}; e^{\frac{2\pi i (\go_1 + \go_2 + 2 \go_4)}{p \go_2 + (p-q) \go_4}}, e^{\frac{2\pi i(\go_3 - \go_4)}{p \go_2+ (p-q) \go_4}})_\infty
   (e^{\frac{2\pi i z}{p \go_2 + (p-q) \go_3}}; e^{\frac{2\pi i (\go_1 + \go_2 + 2 \go_3)}{p \go_2 + (p-q) \go_3}}, e^{\frac{2\pi i(\go_4 - \go_3)}{p \go_2+ (p-q) \go_3}})_\infty ~,\nn\\
   \label{factor_S_3_Ypq}\eea
 where we made use of the identities (\ref{app-relations-1}) and (\ref{app-relations-2}).  For the above formula to be valid we have to assume that
 ${\rm Im} [(\go_1 + \go_2 + 2 \go_{3,4})/(p \go_2 + (p\pm q) \go_{3,4})] \neq 0$ and ${\rm Im} ([(\go_4- \go_3)/(p \go_2 + (p\pm q) \go_{3,4})] \neq 0$.
 The prefactor $B^{Y^{p,q}}$ is given by the following combination of Benoulli polynomials
 \bea
&& B^{Y^{p,q}}(z|\go_1, \go_2, \go_3, \go_4) =
-\frac{i\pi}{6}\Big(B_{3,3}(x| \go_1 +\go_2 + 2\go_3, p\go_2 + (p-q)\go_3, \go_4 - \go_3)\nn\\
&&+B_{3,3}(x| \go_1 +\go_2 + 2\go_4, p\go_2 + (p-q)\go_4, \go_3 - \go_4)\nn\\
&&+B_{3,3}(x| \go_1 + \go_2 + 2 \go_3, p\go_1 +(p+q) \go_3, \go_4 - \go_3)\nn\\
&&+B_{3,3}(x| \go_1 +\go_2 + 2\go_4, p\go_1 + (p+q)\go_4, \go_3 - \go_4)\Big)\nn\\
&&-\frac{i\pi}{2}\Big(B_{2,2}(x| \go_1 +\go_2 + 2\go_4,\go_3 - \go_4)+B_{2,2}(x| \go_1 +\go_2 + 2\go_3, \go_4 - \go_3)\Big)\label{app-gener-bern}.\eea
Using the properties of the Bernoulli polynomials, we can choose to eliminate the $B_{2,2}$'s and get
\bea
&& B^{Y^{p,q}}(z|\go_1, \go_2, \go_3, \go_4) =\nn \\
&& -\frac{i\pi}{6}\sum_{n=0}^{p-1} \Big [
B_{3,3}(z+n(\go_1 +\go_2 + 2\go_3)| p\go_2 + (p-q)\go_3, p\go_1 +(p+q) \go_3 ,\go_4 - \go_3)\nn\\
&&+B_{3,3}(z+n(\go_1 +\go_2 + 2\go_4)| p\go_2 + (p-q)\go_4, p\go_2 + (p+q)\go_4,\go_3 - \go_4) \Big ] ~.\label{Ypq_prefactor_new}
\eea

\providecommand{\href}[2]{#2}\begingroup\raggedright\endgroup

\end{document}